\documentclass[pra,twocolumn,superscriptaddress,showpacs]{revtex4}
\usepackage{graphicx}
\usepackage{epsfig}
\usepackage{bm,hyperref}
\usepackage{CJK}
\usepackage{amsmath}

\begin{document}

\title{Stability of $p$-orbital Bose-Einstein condensates in
  optical checkerboard and square lattices}

\begin{CJK*}{UTF8}{gbsn}
\author{Yong Xu(徐勇)}
\affiliation{Institute of Physics, Chinese Academy of Sciences,
Beijing 100190, China}
\affiliation{International Center for Quantum Materials, Peking University,
Beijing 100871, China}
\author{Zhu Chen(陈竹)}
\affiliation{Institute for Advanced Study, Tsinghua University,  Beijing 100084, China}
\author{Hongwei Xiong (熊宏伟)}
\affiliation{State Key Laboratory of Magnetic Resonance and Atomic and Molecular Physics,
Wuhan Institute of Physics and Mathematics, Chinese Academy of Sciences, Wuhan 430071, China}
\affiliation{Department of Applied Physics, Zhejiang University of Technology, Hangzhou 310023, China}
\author{W. Vincent Liu(刘文胜)}
\affiliation{Department of Physics and Astronomy, University of Pittsburgh, Pittsburgh,
Pennsylvania 15260, USA}
\author{Biao Wu(吴飙)}
\email{wubiao@pku.edu.cn}
\affiliation{International Center for Quantum Materials, Peking University,
Beijing 100871, China}
%
%

%
%
\date{\today}
%
\begin{abstract}
We investigate $p$-orbital Bose-Einstein condensates in both the
square and checkerboard lattice by numerically solving the
Gross-Pitaevskii equation. The periodic potential for the latter
lattice is taken exactly from the recent
experiment [Nature Phys. 7, 147 (2011)]. It is confirmed that the staggered
orbital-current state is the lowest-energy state in the $p$ band. Our
numerical calculation further reveals that for both lattices the
staggered $p$-orbital state suffers Landau instability but the situation
is remarkably different for dynamical instability. A dynamically
stable parameter region is found for the checkerboard lattice, but not
for the square.
\end{abstract}
\pacs{03.75.Kk, 03.75.Lm, 37.10.Jk, 05.30.Jp}
\maketitle
\end{CJK*}
\section{Introduction}
Orbital physics is important in solid-state material due to its
key role in understanding many interesting phenomena including
metal-insulator transition, unconventional superconductivity, and
colossal magnetoresistance~\cite{Tokura-Nagaosa:2000sci}. However, to
fully understand the role of orbital degree of freedom in real solid
materials is challenging because of their complex nature.
A quantum degenerate gas in the optical lattice~\cite{Bloch08RMP,Lim2008PRL},
which is disorder free and highly tunable, is an ideal platform to explore
high orbital physics as the orbital degree of freedom in such an ultracold
gas is separate from spin and charge freedom automatically.
Moreover, a system of neutral bosons loaded into an optical
lattice at low enough temperature has no counterpart in real
quantum materials. Bosonic atoms can condensate into non-ground state,
opening the possibility to explore physics that previously might have
seemed academic or impossible, e.g., the time-reversal symmetry breaking
superfluidity in the nodal $p$ band~\cite{Isacsson2005pra,Liu2006pra,Kuklov:06prl}.

Several experimental methods have been developed to populate the $p$ and higher
orbital bands in
optical lattices~\cite{philips2005pra,Porto:06pra,Anderlini+PhillipsJPB06,Lee+PortoPrl07,
Anderlini+PortoNature07,Muller2007prl,Hemmerich2010Nature,
Olschlarger+Hemmerich:11prl,Olschlarger+Hemmerich:12prl,Soltan+Sengstock:12}.
Pioneering experiments have been carried out by accelerating the
lattice~\cite{philips2005pra}, dynamically deforming the double-well
potentials as a single site manipulation~\cite{Porto:06pra,Anderlini+PhillipsJPB06,
Lee+PortoPrl07,Anderlini+PortoNature07}, and
exciting atoms into the higher vibrational state along a controlled
lattice direction through stimulated Raman
transitions~\cite{Muller2007prl}. The recent implementation of orbital
degrees of freedom in
checkerboard~\cite{Hemmerich2010Nature,Olschlarger+Hemmerich:11prl,Olschlarger+Hemmerich:12prl}
and hexagonal~\cite{Soltan+Sengstock:12} optical lattices truly opens
an era of exploring orbital phases of quantum matter that have no
prior analogues in solid state materials. For example, the experiments
of \"Olschl\"arger
\emph{et al}~\cite{Hemmerich2010Nature,Olschlarger+Hemmerich:11prl,Olschlarger+Hemmerich:12prl}
show that bosonic atoms are loaded and kept in the excited $p$-orbital bands for nearly as
long as the ultracold gases can be, thus effectively possessing infinite long
lifetime in the scale of tunneling.
In the experiment, atoms are transferred from the $s$-orbital
band to the $p$-orbital band through the changing of the double-well relative depth,
the time of flight images have illustrated the  macroscopic  occupation in the $p$-orbital
real state and complex state. The same group also reported the observation of the
exotic $d$- and $f$-band superfluid phases~\cite{Olschlarger+Hemmerich:11prl, Olschlarger+Hemmerich:12prl}.
Dynamically, fermionic and bosonic atoms are made across from lower to
high orbital bands in optical honeycomb~\cite{Tarruell+Esslinger:12nat} and
checkerboard~\cite{Olschlarger+Hemmerich:12prl} lattices, respectively.

Although the experiments are done with continuous optical potential of
periodic oscillation, most of the past theoretical work employed the standard
tight-binding method and approximated the system to a lattice model, where the
atoms are strictly constrained to the $p$-orbital. Many interesting results
have been worked out, such as the staggered state as the ground state in the
$p$-orbital band~\cite{Liu2006pra},
superfluid transition to Mott phase~\cite{Isacsson2005pra,Collin+MartikainenPRA10,Li+WVLPRA11},
supersolid quantum phase~\cite{DasSarma2005PRL} and quantum strip ordering
in triangular lattices~\cite{Wu2006PRLT} (for other interesting studies and a brief perspective,
see, for example, Ref.~\cite{Liu2011Nature}). In this work, we use the continuous
model where little theoretical work has been done besides a variational
computation of the lowest energy in the $p$-orbital band~\cite{Cai2011pra}.  The
continuous model provides a better and complete description of the system
as it can capture the decay to the $s$ band that happens in real experiments and
applies beyond tight-binding approximation. We focus on the stability of the
$p$-orbital state, which is crucial to understand the $p$-band superfluidity. To
understand the superfluidity in the $s$ band, stability of the Bloch states in the
$s$ band has been analyzed theoretically for different lattices---one-dimensional
lattices~\cite{Wu2001pra}, two-dimensional square lattices~\cite{wu2010PRA},
and two-dimensional double-well lattices~\cite{HHui+Porto2012arXiv}. It was
also investigated experimentally for a one-dimensional lattice~\cite{Fallani2004prl}.
However, for the $p$-orbital
system, to the best of our knowledge,
only one paper~\cite{Martikainen2011pra}
discussed the dynamical instability and it is limited
to the one-dimensional case.

In this paper, with the Gross-Pitaevskii (GP) equation, we calculate exactly the
$p$-orbital band ground state for both the two-dimensional (2D) square optical
lattice and the checkerboard lattice used in the
experiment~\cite{Hemmerich2010Nature}. We confirm that the lowest-energy state
in the $p$-band is the staggered state found in Ref.~\cite{Liu2006pra}. The Landau
instability and dynamical instability of this state are investigated. For
the lattice model approach, such a study of stability is not accurate because
the $s$ band is always removed from the Hamiltonian to make the condensate
impossible to decay. Our calculation shows that, for both periodic potentials,
the staggered state always has Landau instability as the state is a local saddle
point that can decay into the $s$ band. For the dynamical stability, these two 2D
lattices are very different: our numerical search does not find any parameter region
for the square lattice, where the staggered state is dynamically stable; in contrast, there
exists a parameter region for the checkerboard potential where the staggered
state is dynamically stable. This is consistent with the intuitive
understanding that the checkerboard potential offers better
stability~\cite{Stojanovi2008prl}. That is, on general ground, the checkerboard
potential may be viewed as a particular configuration of the simple
double-well lattice potential, and the energy gap between the lowest $s$ and the first
excited $p$ orbital bands is much smaller than that between the $p$ and the higher
excited bands. Consequently, the first-order decay of atoms in the $p$ band  due to
scattering for
the checkerboard lattice is suppressed by energy conservation according to Fermi
golden rule, in contrast with the decay for
the square lattice of single wells where band spacings are approximately equal in the
tight-binding (i.e., the simple harmonic oscillator) limit.

The paper is organized in the following way. In Sec. \ref{BE},
the general theoretical framework of our calculation is given.
In Secs. \ref{SL} and \ref{SA}, the results for both the square
lattice and the checkerboard lattice are presented, respectively.
Finally, conclusions are drawn in Sec. \ref{CL}.

\section{General theoretical framework}
\label{BE}
We consider the Bose-Einstein condensate of bosons in a 2D optical potential
with periodicity characterized by two lattice vectors to be defined below.
To compare with realistic three-dimensional experimental  systems, our model
applies to the experiments where a strong trap is applied along the third
direction. We thus neglect the third dimension, which only contributes to
the effective interaction parameter. The 2D GP equation is
\begin{equation}
\label{G_tGP}
i\hbar\partial_{t}\psi({\rm \bf r})=\Big[-\frac{\hbar^2}{2m}\nabla^2+V({\rm \bf r})+g|\psi|^2\Big]\psi({\rm \bf r}),
\end{equation}
where $V({\rm \bf r}+{\rm \bf a}_1)=V({\rm \bf r}+{\rm \bf a}_2)=V({\rm \bf r})$ with ${\rm \bf a}_1$ and ${\rm \bf a}_2$ the lattice vectors;
$m$ is the atomic mass, and $g$ is the interaction parameter. $\psi({\rm \bf r})$ is normalized as
$\frac{1}{\Omega}\int_{\Omega}{|\psi({\rm \bf r})|^2d{\rm \bf r}}=1$ where the subscript $\Omega$
indicates an integral over the unit cell with an area $\Omega=|{\rm \bf a}_1||{\rm \bf a}_2|$.

We are interested in the lowest-energy state in the $p$-orbital band. This type of state
must be stationary and satisfy the time-independent GP equation
\begin{equation}
\label{G_GP}
\Big[-\frac{\hbar^2}{2m}\nabla^2+V({\rm \bf r})+g|\psi|^2\Big]\psi({\rm \bf r})=\mu\psi({\rm \bf r}),
\end{equation}
where $\mu$ is the chemical potential. The extended solution to the above nonlinear
periodic equation  has the form $\psi({\rm \bf r})=e^{i{\rm \bf k}\cdot{\rm \bf r}}f({\rm \bf r})$. For the usual Bloch states,
$f$ is a period function with the same period as that of the optical lattice, $f({\rm \bf r}+{\rm \bf a}_1)=f({\rm \bf r})$ and
$f({\rm \bf r}+{\rm \bf a}_2)=f({\rm \bf r})$. Besides Bloch states, there are other solutions such as the period-doubled
solutions~\cite{machholm2004PRA,wu2010PRA},  where $f$ satisfies
$f({\rm \bf r}+2{\rm \bf a}_1)=f({\rm \bf r})$ and $f({\rm \bf r}+2{\rm \bf a}_2)=f({\rm \bf r})$. For the $s$ band, the usual Bloch states always
have lower energy than that of period-doubled states. But for the $p$-orbital band, previous
studies~\cite{Liu2006pra, Cai2011pra} have shown that the period-doubled solution
has lower energy than the corresponding Bloch state due to the extra $\pi$ phase
in $p$-orbital tunneling.

For the two types of lattices considered in this work, the following two Bloch states
are degenerate and have the lowest-energy among all the Bloch states,
\begin{equation}
\label{px}
P_{x}=\sum_{{\rm \bf G}}u_{{\rm \bf G}}e^{i({\rm \bf k}_1+{\rm \bf G})\cdot{\rm \bf r}}\,,
\end{equation}
and
\begin{equation}
\label{py}
P_{y}=\sum_{{\rm \bf G}}v_{{\rm \bf G}}e^{i({\rm \bf k}_2+{\rm \bf G})\cdot{\rm \bf r}},
\end{equation}
where  ${\rm \bf G}=m{\rm \bf b}_1+n{\rm \bf b}_2$, ${\rm \bf b}_1$ and ${\rm \bf b}_2$ are reciprocal lattice vectors, and
  ${\rm \bf k}_1={\rm \bf b}_1/2$ and ${\rm \bf k}_2={\rm \bf b}_2/2$. Substituting these two equations into Eq.(\ref{G_GP})
leads to a series of nonlinear equations of either $u$ or $v$.
We use the subroutine fsolve of MATLAB to solve these equations.

There are other types of solutions, which can be symbolically expressed
as either
 \begin{equation}
 \label{DP_sL1}
 P_{x\pm y}=\frac{1}{\sqrt{2}}(P_x\pm P_y)\,,
 \end{equation}
 or
 \begin{equation}
 \label{DP_sL2}
 P_{x\pm iy}=\frac{1}{\sqrt{2}}(P_x\pm iP_y)\,.
 \end{equation}
These solutions are period-doubled states and therefore non-Bloch. Without interaction, these
non-Bloch states would have the same energy as the Bloch states $P_x$ and $P_y$.
With interaction, these states may break the degeneracy, splitting into either lower
or higher energy. To find these non-Bloch states, one can
similarly substitute Eqs. (\ref{DP_sL1}) and (\ref{DP_sL2}) into Eq. (\ref{G_GP}) and find the
coefficients $u$ and $v$ numerically. The coefficients $u$ and $v$ found
here are in general different from the ones found by substituting Eqs. (\ref{px}) and (\ref{py}) into
Eq. (\ref{G_GP}). This is the essential technical difference from the work in Ref.~\cite{Cai2011pra}.
Due to the time-reversal symmetry, it is sufficient to consider only $P_{x+y}$ and $P_{x+iy}$.

We are interested in the stability of these $p$-orbital states. We know that
the lowest-energy state in the $s$ band is always stable because it is the 
lowest-energy state of the system. It is imperative to know whether the lowest-energy
$p$-orbital state is stable or not. In fact, this was already the concern at the beginning
of studying the $p$-orbital states in cold-atom systems~\cite{Liu2006pra,Kuklov:06prl}
as the decay to the $s$ band seems almost inevitable. However, it is possible that
interaction may be able to stabilize a certain $p$-orbital state and make it a metastable state.
The primary purpose of this work is to answer whether such a possibility exists.
As will be shown later, the state $P_{x+iy}$
always has the lowest-energy among all examined $p$-orbital states. Consequently, we
will focus on the stability of this state.

To examine the stabilities of a state, one can follow the well-known
procedure~\cite{Wu2001pra} and obtain the Bogoliubov equation in momentum ${\rm \bf q}$ space
\begin{equation}
\label{G_BdG}
\epsilon_{{\rm \bf q}}\begin{pmatrix}u_{{\rm \bf q}} \\ v_{{\rm \bf q}}\end{pmatrix}=\sigma_zM\begin{pmatrix}u_{{\rm \bf q}} \\ v_{{\rm \bf q}}\end{pmatrix}\,,
\end{equation}
where $\sigma_z$ is the Pauli matrix.  For the state $P_{x+iy}$, we have
\begin{equation}
M=\begin{pmatrix} \mathcal{L}({\rm \bf q}) & gP^2_{x+iy}\\ gP^{*2}_{x+iy}& \mathcal{L}({\rm \bf q})\end{pmatrix},
\end{equation}
with
\begin{eqnarray}
\mathcal{L}({\rm \bf q})=-\frac{\hbar^2}{2m}[(\partial_x+iq_x)^2+(\partial_y+iq_y)^2] \nonumber \\
+V({\rm \bf r})-\mu_{xy}+2g|P_{x+iy}|^2.
\end{eqnarray}
Since $P_{x+iy}$ is period-doubled, we have $u_{\rm \bf q}({\rm \bf r}+2{\rm \bf a}_1)=u_{\rm \bf q}({\rm \bf r}+2{\rm \bf a}_2)=u_{\rm \bf q}({\rm \bf r})$ and
 $v_{\rm \bf q}({\rm \bf r}+2{\rm \bf a}_1)=v_{\rm \bf q}({\rm \bf r}+2{\rm \bf a}_2)=v_{\rm \bf q}({\rm \bf r})$. To numerically diagonalize the matrix $\sigma_zM$,
we expand $u$ and $v$ in Fourier series as $u_{{\rm \bf q}}({\rm \bf r})=\sum_{{\rm \bf G}}u_{{\rm \bf G}}e^{i{\rm \bf G}\cdot{\rm \bf r}/2}$
and $v_{{\rm \bf q}}({\rm \bf r})=\sum_{{\rm \bf G}}v_{{\rm \bf G}}e^{i{\rm \bf G}\cdot{\rm \bf r}/2}$.  The diagonalization of $\sigma_zM$
for the phonon modes
yields the Bogoliubov excitation of the state $P_{x+iy}$. This state has Landau instability
if part of its Bogoliubov excitations is negative;  it has dynamical instability
if part of its Bogoliubov excitations is imaginary~\cite{wu2010PRA}.

\section{Square Lattice}
\label{SL}
The square lattice can be formed by simply overlapping two counter-propagating
laser beams. Mathematically, it is described by
$V(x,y)=V_0\Big[\cos(x)+\cos(y)\Big]$. For this lattice, it is convenient
to use the following time independent GP equation
 \begin{equation}
\label{S_GP}
\Big[-\frac{1}{2}(\partial^2_x+\partial^2_y)+V(x,y)+c|\psi|^2\Big]\psi(x,y)=\mu\psi(x,y).
\end{equation}
The above equation has been made dimensionless by scaling energy with $8E_r$
and length with $1/2k_L$. In this section, $x$ and $y$ are dimensionless.
$E_r$ is the recoil energy and $k_L=2\pi/\lambda$ is
the wave vector of the laser beam. The interaction constant is
$c=mng/\hbar^2$ with $m$
being the atom mass, $n$ the BEC density (the average particle number per site), and
$g=2\sqrt{2\pi}\hbar^2a_s/(\sigma m)$, where $a_s$ is the
$s$-wave scattering length and $\sigma$ is the characteristic length of the harmonic
trap along the $z$ direction.

\begin{figure}[!ht]
\scalebox{0.43}[0.43]{\includegraphics[3,371][572,640]{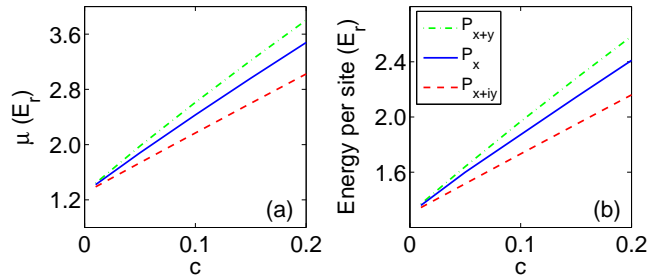}}
\caption{(Color online) (Square lattice) Chemical potential (left) and energy per lattice site (right) for
the $P_{x}$ state (blue solid line), $P_{x+iy}$ state (red dashed one)
and $P_{x+y}$ (green dashed dot line) in the square lattice. $V_0=0.8E_r$.}
\label{Squar_mu_c}
\end{figure}

Following the procedure described in the above section, we have
numerically computed three states $P_x$, $P_{x+y}$, and $P_{x+iy}$.
Fig. \ref{Squar_mu_c} illustrates how the energies and chemical potentials
change with the interaction constant $c$ for these three states. It is clear
from the figure that state $P_{x+iy}$ always has the lowest-energy
and the energy gap to the other states increases with $c$. This confirms
the earlier results obtained with lattice model~\cite{Liu2006pra} and variational
method~\cite{Cai2011pra}. It is also shown in Fig.~\ref{Squar_mu_c} that
state $P_{x+y}$ always has the highest energy among the three.

\begin{figure}[!ht]
\begin{center}
\scalebox{0.53}[0.53]{\includegraphics[0,0][465,495]{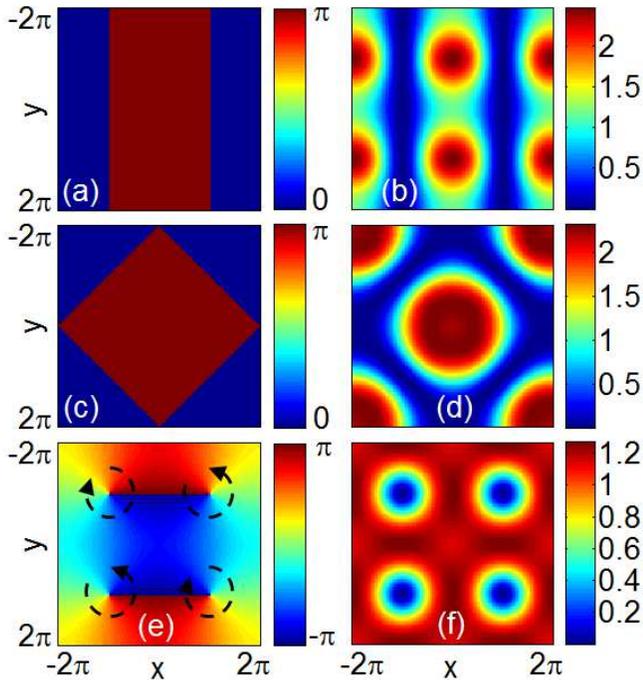}}
\caption{(Color online) (Square lattice) Phase and density profile for state $P_x$ (a), (b),
state $P_{x+y}$ (c), (d), and state $P_{x+iy}$ (e), (f), respectively, for the square lattice.
Arrows in (e) indicate the vortex rotating directions. For phase in (a) and (c),
the dark blue is $0$, the brown is $\pi$. $x$ and $y$ have units of $1/2k_L$.
$V_0=0.8E_r$ and $c=0.01$.}
\label{2D_phase}
\end{center}
\end{figure}

The phase and density profiles of these three states are shown in Fig. \ref{2D_phase}.
These three states not only differ in phase but also in density. Since the wave
functions for both states $P_x$ and $P_{x+y}$ are real,
their phase can only be either $0$ or $\pi$.
Specifically, the state $P_x$ has a stripe phase structure while the state $P_{x+y}$
has a square-shaped one. The wave function of the state $P_{x+iy}$ is complex
and breaks time-reversal symmetry. Consequently, this state has much richer
phase structure, which is evidently shown by the staggered orbital currents in
Fig. \ref{2D_phase}(e). This feature of staggered orbital currents is the most
prominent predication in Ref.~\cite{Liu2006pra}.



\begin{figure}[!ht]
\scalebox{0.4}[0.4]{\includegraphics[60,0][535,340]{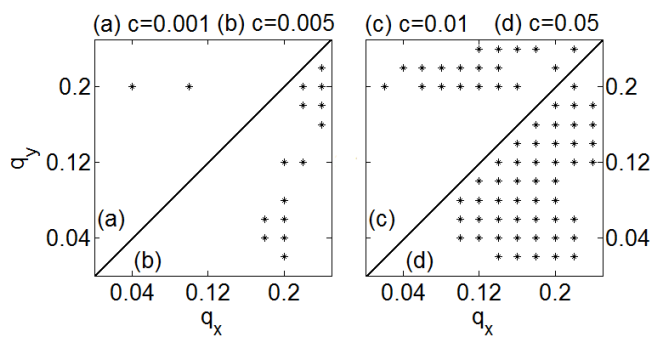}}
\caption{(Square lattice) Dynamical stability diagram of state $P_{x+iy}$
for the square lattice at $V_0=0.8E_r$. The region of dynamical instability
is marked by stars. $q_x$ and $q_y$ are in unites of $2k_L$ with a step length $0.02$.
Since the diagram is symmetric with respect to $q_x$ and $q_y$, we only
draw the upper triangle for (a) and (c) and the lower one for (b) and (d).
The upper limit for $q_x$ and $q_y$ is $0.25$ is  due to that $P_{x+iy}$ is a
period-doubled state.
}
\label{Squar_sta}
\end{figure}

As the state $P_{x+iy}$ has the lowest-energy among the $p$-orbital states, we focus on the stability
of this state. It is examined through its Bogoliubov excitations by diagonalizing the matrix
$\sigma_zM$. Our computation finds that the Bogoliubov excitations always have a negative part,
indicating that the $P_{x+iy}$ has Landau instability and is not a metastable state.
The situation is more delicate  for dynamical
stability. Figure \ref{Squar_sta} shows the phase diagram of dynamical instability, where
the stars mark out the region of the (momentum) ${\rm \bf q}$ space where the Bogoliubov excitations are imaginary.
It is clear from the figure that the stable region of the ${\rm \bf q}$ space increases
as the interaction $c$ decreases. It is reasonable to expect that the whole region be stable
when $c$ is small enough. However, within our numerical capability, we are not able
to identify the values of $c$ and $V_0$ for which the state $P_{x+iy}$ is free of dynamical
instability. As shown in Figs. \ref{Squar_sta}(a) and \ref{Squar_sta}(b),
even for very small $c$, there are some regions where the excitations are imaginary.
This means that state $P_{x+iy}$ is  dynamically stable only for extremely small values of $c$.
We have attempted to calculate the critical $c$ for the typical experimental $V_0$~\cite{Bloch08RMP}
and found that they are of the order of $10^{-4}$. However, despite of intensive efforts,
our numerical method is in capable of pinning down the exact value of these critical
$c$ as indicated by the irregular black line in the inset of Fig. \ref{st_phase_dia},
where the parameter region of dynamical instability for the square lattice is marked out.
These results imply that it is almost impossible to use the square lattice to study $p$-orbital BEC
states experimentally as dynamical instability can destroy a BEC
in tens of milliseconds~\cite{Fallani2004prl}.

\section{checkerboard potential}
\label{CP}
\label{SA}
The optical lattice used in the experiment~\cite{Hemmerich2010Nature} is
a checkerboard potential described by
\begin{eqnarray}
V(x,y)&&=-\frac{V_0}{4}\Big|\eta\Big[\big({\rm \bf e}_z\cos(\alpha)+{\rm \bf e}_y\sin(\alpha)\big)e^{ik_L x}+\nonumber\\
&&\epsilon{\rm \bf e}_ze^{-ik_L x}\Big]+e^{i\theta}{\rm \bf e}_z\big(e^{ik_L y}+
\epsilon{\rm \bf e}^{-ik_L y}\big)\Big|^2\,,
\end{eqnarray}
where ${\rm \bf e}_y$ and ${\rm \bf e}_z$ are unit vectors in each direction.
Here $x$ and $y$ are the space coordinates,
$k_L$ is the laser wave vector, $\alpha$ is the polarization angle to the $z$ direction,
$\epsilon$ is the reflection loss, $\eta$ describes the small power difference
between two interferometers,  $\theta$ is the phase difference between the
beams propagating in the $x$ and $y$ directions,  and $V_0$ is determined by the
laser power. The angle $\alpha$ can be used to adjust the degree of anisotropy:
when $\alpha=\pi/5$, the energy minimum points of the two $p$-orbital Bloch
bands are degenerate. The phase difference $\theta$
controls the relative depth of a double-well. In the experiment, bosonic atoms are loaded
to the $p$-orbital band by adjusting $\theta$, $\eta\approx0.95$, $\epsilon\approx0.81$,
and $V_0=6.2E_r$ with the recoil energy $E_r=\hbar^2k_L^2/2m$.

To have a dimensionless time-independent GP equation as in Eq. (\ref{S_GP}), we
scale energy with $4E_r$ and length with $1/\sqrt{2}k_L$. In the dimensionless
expression, the lattice vectors of the potential are
${\rm \bf a}_1=\sqrt{2}\pi({\rm \bf e}_x+{\rm \bf e}_y)$, ${\rm \bf a}_2=\sqrt{2}\pi(-{\rm \bf e}_x+{\rm \bf e}_y)$ and reciprocal
vectors are ${\rm \bf b}_1=({\rm \bf e}_x+{\rm \bf e}_y)/\sqrt{2}$ and ${\rm \bf b}_2=(-{\rm \bf e}_x+{\rm \bf e}_y)/\sqrt{2}$. The
quasi-momentum and coordinate vector are, ${\rm \bf k}=k_x{\rm \bf b}_1+k_y{\rm \bf b}_2$ and
${\rm \bf r}=(x{\rm \bf a}_1+y{\rm \bf a}_2)/2\pi$, respectively.

\begin{figure}[!ht]
\scalebox{0.43}[0.43]{\includegraphics[2,355][575,617]{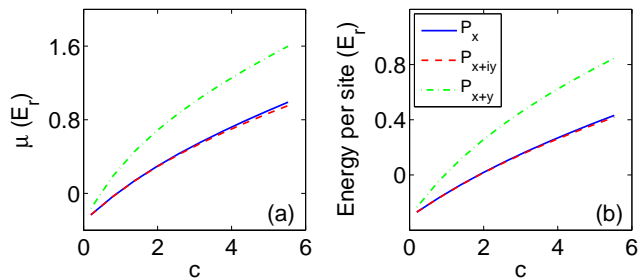}}
\caption{(Color online) (Checkerboard lattice) Chemical potential (left)
and energy per lattice well (right) for the $P_{x}$ state (blue solid line),
$P_{x+iy}$ state (red dashed one), and $P_{x+y}$ (green dashed dot one)
in the checkerboard potential. $V_0=6.2E_r$.}
\label{Exp_mu_c}
\end{figure}

We have done  the same set of numerical computation for
the three states $P_x$, $P_{x+y}$, and $P_{x+iy}$ in the checkerboard lattice
as in the square lattice. Figure \ref{Exp_mu_c} shows their energies and
chemical potentials as a function of the interaction constant $c$. Similar to
the square lattice, the state $P_{x+iy}$ in the checkerboard lattice is found
to have the lowest-energy too. There is however an evident difference, i.e.,
for the checkerboard, the states $P_{x+iy}$ and $P_x$ are very close in energy
while state $P_{x+y}$ has much higher energy. This feature suggests that the
state observed in the experiment~\cite{Hemmerich2010Nature} is probably not
$P_{x+y}$.

\begin{figure}[!ht]
\begin{center}
\scalebox{0.42}[0.42]{\includegraphics[5,5][620,655]{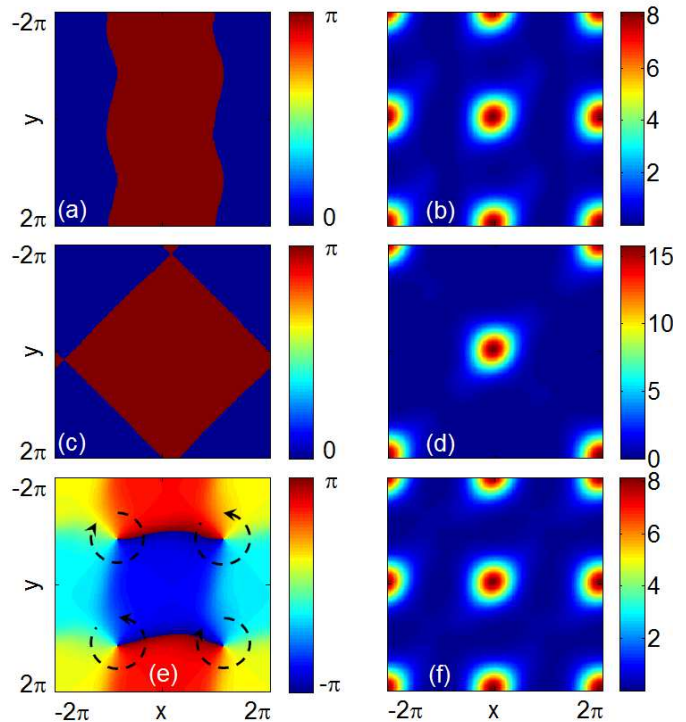}}
\caption{(Color online) (Checkerboard lattice) Phase and density profile for state $P_x$ (a), (b),
state $P_{x+y}$ (c), (d)
and state $P_{x+iy}$ (e), (f), respectively, in the checkerboard potential.
Arrows in (e) indicate vortex rotating directions.
In (a) and (c), the dark blue is for phase of $0$ and the brown for phase of $\pi$.
$x$, and $y$ have units of $1/\sqrt{2}k_L$. $V_0=6.2E_r$, $c=0.2$. }
\label{Exp_phase}
\end{center}
\end{figure}

The phase and density profiles of these three states are illustrated in
Fig.~\ref{Exp_phase}. The phase profiles show a structure similar to that for
the square lattice, such that $P_x$ has the wavelike stripe phase structure,
$P_{x+y}$ has the square phase structure and $P_{x+iy}$ has the staggered
orbital currents structure~\cite{Liu2006pra}. This result supports the
  conclusion that one may use the square lattice as a simplified theoretical
  model to understand much of the unconventional properties of the $p$-orbital
  condensates as observed in the more complex, experimentally realized
  checkerboard lattice. In terms of dynamical instability, the two lattice
  configurations are however qualitatively different, to be elaborated below.
The density profiles do not show clear difference between state $P_x$ and
$P_{x+iy}$ and the reason is that the probability density in the deeper well
where the vortex appears is very small.

\begin{figure}[!ht]
\scalebox{0.41}[0.41]{\includegraphics[80,0][535,340]{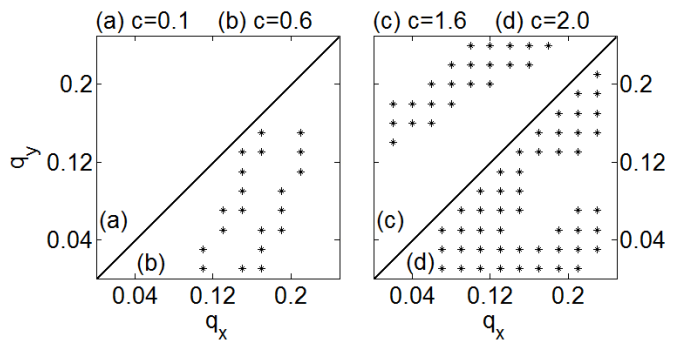}}
\caption{ (Checkerboard lattice) Dynamical stability diagram of the state
$P_{x+iy}$ of the checkerboard
potential. The region of dynamical instability is marked by stars. The blank panel in (a) indicates
that the system is free of dynamical instability for $c=0.1$, to be contracted with that
for the square lattice in Fig.~\ref{Squar_sta}.
$q_x$ and $q_y$ are in unites of $\sqrt{2}k_L$ with a step length $0.02$. $V_0=6.2E_r$.
Since the diagram is symmetric with respect to $q_x$ and $q_y$, we only
draw the upper triangle for (a) and (c) and the lower one for (b) and (d).
The upper limit for $q_x$ and $q_y$ is $0.25$ is  due to that $P_{x+iy}$ is a
period-doubled state. }
\label{s_exp}
\end{figure}

\begin{figure}[!ht]
\scalebox{0.5}[0.5]{\includegraphics[130,0][349,430]{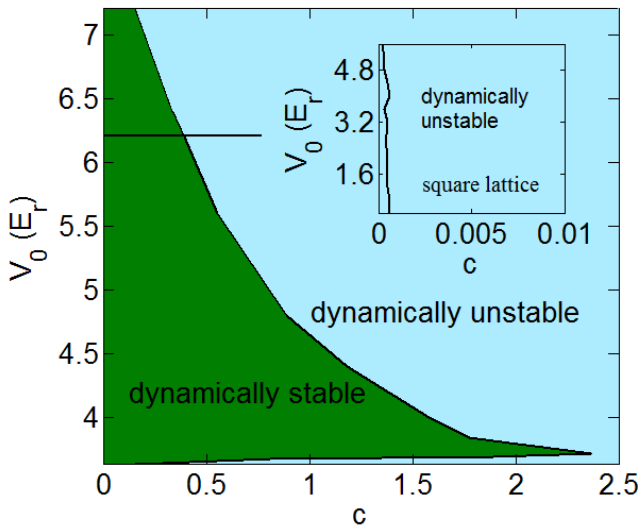}}
\caption{(Color online) (Checkerboard lattice) Stable and unstable region in the system
parameter space for state $P_{x+iy}$ in the checkerboard potential.
The black line indicates the parameter range used in
the experiment~\cite{Hemmerich2010Nature} where $V_0=6.2E_r$.
The inset shows the stability regions for the square lattice.
The irregularity of the solid line in the inset is caused by the inability of  our numerical method
to compute precisely the critical value of $c$ below which the system is dynamically stable.}
\label{st_phase_dia}
\end{figure}

For stability, we focus on that of state $P_{x+iy}$ just as in the square
lattice. We investigate it also through Bogoliubov excitations
by diagonalizing the matrix $\sigma_zM$. Similar to the square lattice, our
calculation shows that the Bogoliubov excitations in the checkerboard
always have negative part, indicating that state $P_{x+iy}$ in the experiment
also has Landau instability. For dynamical stability, the unstable region in
the ${\rm \bf q}$ space increases with $c$ as shown in
Fig.~\ref{s_exp}. However, there is a crucial difference from the case of
square lattice: there exists a critical value of $c$, below which there
is no dynamical instability as indicated by the blank panel in Fig.~\ref{s_exp}(a).
We are able to mark out a region in the space
spanned by the system parameters $c$ and $V_0/E_r$, where
the $P_{x+iy}$ state is free of
dynamical instability (shown in the phase diagram Fig.~\ref{st_phase_dia}).
Due to the uncertainty of the BEC density,
we have marked the experimental parameter range~\cite{Hemmerich2010Nature} with
a solid line.  This shows that it is likely that the $P_{x+iy}$ for the
experimental setup is dynamically stable.  Since the time scale for Landau
instability is of the order of 500 ms~\cite{Fallani2004prl} which is much longer
than that of the experiment~\cite{Hemmerich2010Nature}, it is reasonable that
Landau instability does not have much effect.

In order to make sure that our calculation is correct, we simulate the real
system of Eq.(\ref{G_tGP}) with the split-operator method. We evolve numerically
a BEC in the $P_{x+iy}$ state with a small perturbation $\delta\psi$ ($10\%$).
When $c=0.2$, the simulation shows that the state is stable.
When $c=3.0$ and $c=7.9$, the simulation
shows that the state is destroyed after $t=17.5$ms and $t=2.5$ms, respectively.
All results in the three simulations are consistent with our Bogoliubov excitation
calculation.

To map out the phase diagram of dynamical instability in Fig.~\ref{st_phase_dia}
experimentally, one may need to use the Feshbach resonance to tune the interaction
strength. When the Feshbach resonance is not available, one can still observe the
effect of dynamical instability by turning up the laser power to drive the system
into dynamically unstable regime. The effects of dynamical instability should be
similar to what was observed in Ref.~\cite{Fallani2004prl}.

\section{Conclusion}
\label{CL}

In conclusion, we have examined a cold gas of  interacting bosonic atoms
loaded in two optical lattice geometries, namely,  the square and the
checkerboard lattices,  for which unconventional $p$-orbital Bose-Einstein
condensates have been under active investigation in recent years, both
theoretically and experimentally.  The usual theoretical approach
used in the past is to assume the standard tight-binding approximation and
conveniently reduce the system to a Hubbard-like lattice model of one single
orbital band of interest, i.e., the $p$ band. The present approach is however
different. Here, the model system is solved numerically with the GP
equation of microscopic two-body interaction by treating the optical lattice
exactly as a continuous, periodic potential, in which both the ground state
$s$-band and all the higher orbital bands are included. The approach thus
is capable of providing a complete analysis for the Landau and dynamical
instabilities  of a $p$-orbital BEC. Such a complete analysis for instability
was not considered before, to the best of our knowledge. We find that
the staggered state $P_{x+iy}$ indeed has the lowest-energy in the $p$ band.
By computing the Bogoliubov excitation, we further find that for both
lattices Landau instability is present, which shows that the staggered state
is not really a state at local energy minimum. For dynamical
stability, we find that there exists a parameter region where the
staggered state is free of dynamical instability for the checkerboard lattice
whereas no such a parameter region is found for the square lattice. This
suggests that the staggered state be of long life time in the former, but
not the latter.

\acknowledgments The authors are grateful for very helpful discussions with
A. Hemmerich. Y.X., Z.C., and B.W. are supported by the
National Basic Research Program of MOST (NBRP) of China
(2012CB921300, 2013CB921900), the National Natural Science
Foundation (NSF) of China (10825417, 11274024), the
Research Fund for the Doctoral Program of Higher Education
(RFDP) of China (20110001110091). W.V.L is supported by
the US DOD AFOSR (FA9550-12-1-0079), ARO (W911NF-
11-1-0230),DARPAOLEProgram throughARO, and theNSF
of China (11128407).H.W. is supported by the NBRP of China
(2011CB921503) and the NSF of China (11175246).


\begin{thebibliography}{25}
\expandafter\ifx\csname natexlab\endcsname\relax\def\natexlab#1{#1}\fi
\expandafter\ifx\csname bibnamefont\endcsname\relax
  \def\bibnamefont#1{#1}\fi
\expandafter\ifx\csname bibfnamefont\endcsname\relax
  \def\bibfnamefont#1{#1}\fi
\expandafter\ifx\csname citenamefont\endcsname\relax
  \def\citenamefont#1{#1}\fi
\expandafter\ifx\csname url\endcsname\relax
  \def\url#1{\texttt{#1}}\fi
\expandafter\ifx\csname urlprefix\endcsname\relax\def\urlprefix{URL }\fi
\providecommand{\bibinfo}[2]{#2}
\providecommand{\eprint}[2][]{\url{#2}}





\bibitem{Tokura-Nagaosa:2000sci}{Y. Tokura, and N. Nagaosa, Science
  {\bf 288}, 462 (2000).}
\bibitem{Bloch08RMP} I. Bloch, J. Dalibard, and W. Zwerger, Rev. Mod. Phys. {\bf 80}, 885 (2008).
\bibitem{Lim2008PRL} L. K. Lim, C. M. Smith, and A. Hemmerich, Phys. Rev. Lett. {\bf 100}, 130402 (2008).
\bibitem{Isacsson2005pra}{A. Isacsson, and S. M. Girvin, Phys. Rev. A {\bf 72}, 053604 (2005).}
\bibitem{Liu2006pra}{W. V. Liu, and C. Wu, Phys. Rev. A {\bf 74}, 013607 (2006).}
\bibitem{Kuklov:06prl}{A. B. Kuklov, Phys. Rev. Lett. {\bf 97}, 110405 (2006).}
\bibitem{philips2005pra} {A. Browaeys, H. H\"affner, C. McKenzie, S. L. Rolston,
K. Helmerson, and W. D. Phillips, Phys. Rev. A {\bf 72}, 053605 (2005).}
\bibitem{Porto:06pra}{J. Sebby-Strabley, M. Anderlini, P. S. Jessen, and J. V. Porto,
Phys. Rev. A {\bf 73}, 033605 (2006).}
\bibitem{Anderlini+PhillipsJPB06}{M. Anderlini, J. Sebby-Strabley, J. Kruse,
J. V Porto, and W. D Phillips, J. Phys. B {\bf 39}, S199 (2006).}
\bibitem{Lee+PortoPrl07} {P. J. Lee, M. Anderlini, B. L. Brown, J. Sebby-Strabley,
W. D. Phillips, and J. V. Porto, Phys. Rev. Lett. {\bf 99}, 020402 (2007).}
\bibitem{Anderlini+PortoNature07} {M. Anderlini, P. J. Lee, B. L. Brown, J. Sebby-Strabley,
W. D. Phillips, and J. V. Porto, Nature {\bf 448}, 452 (2007).}
\bibitem{Muller2007prl} {T. M\"uller, S. F\"olling, A. Widera, and I. Bloch,
 Phys. Rev. Lett {\bf 99}, 200405 (2007).}
\bibitem{Hemmerich2010Nature}{G. Wirth, M. \"Olschl\"ager, and A. Hemmerich,
Nature Physics {\bf 7}, 147 (2011).}
\bibitem{Olschlarger+Hemmerich:11prl}{M. \"Olschl\"ager, G. Wirth, and
  A. Hemmerich,
Phys. Rev. Lett. {\bf 106}, 015302 (2011).}
\bibitem{Olschlarger+Hemmerich:12prl}{M. \"Olschl\"ager, G. Wirth,
  T. Kock, and A. Hemmerich,
Phys. Rev. Lett. {\bf 108}, 075302 (2012).}
\bibitem{Soltan+Sengstock:12}{P. Soltan-Panahi, D. -S. L\"uhmann,
  J. Struck, P. Windpassinger, and K. Sengstock,
Nature Physics {\bf 8}, 71 (2012).}
\bibitem{Tarruell+Esslinger:12nat}{L. Tarruell, D. Greif, T. Uehlinger,
G. Jotzu, and T. Esslinger, Nature {\bf 483}, 302
  (2012).}
\bibitem{Collin+MartikainenPRA10} {A. Collin, J. Larson,
      and J. -P. Martikainen, Phys. Rev. A {\bf 81}, 023605 (2010).}
\bibitem{Li+WVLPRA11} {X. Li, E. Zhao, and W. V. Liu, Phys. Rev. A {\bf 83}, 063626 (2011).}
\bibitem{DasSarma2005PRL}{V. W. Scarola, and S. Das Sarma, Phys. Rev. Lett {\bf 95}, 033003 (2005).}
\bibitem{Wu2006PRLT}{C. Wu, W. V. Liu, J. Moore, and S. Das Sarma, Phys. Rev. Lett {\bf 97}, 190406 (2006).}
\bibitem{Liu2011Nature}{M. Lewenstein, and W. V. Liu, Nature Physics {\bf 7}, 101 (2011).}
\bibitem{Cai2011pra}{Z. Cai, and C. Wu, Phys. Rev. A {\bf 84}, 033635 (2011).}
\bibitem{Wu2001pra}{B. Wu, and Q. Niu, Phys. Rev. A {\bf 64}, 061603 (2001).}
\bibitem{wu2010PRA}{Z. Chen, and B. Wu, Phys. Rev. A {\bf 81}, 043611 (2010).}
\bibitem{HHui+Porto2012arXiv}{H. -Y Hui, R. Barnett, J. V. Porto, and S. Das Sarma, arXiv:1208.6300v1 (2012).}
\bibitem{Fallani2004prl}{L. Fallani, L. De Sarlo, J. E. Lye, M. Modugno, R. Saers, C. Fort, and M. Inguscio,
Phy. Rev. Lett {\bf 93}, 140406 (2004).}
\bibitem{Martikainen2011pra} {J. -P. Martikainen, Phys. Rev. A {\bf 83}, 013610 (2011).}
\bibitem{Stojanovi2008prl} {V. M. Stojanovi\'c, C. Wu, W. V. Liu, and S. Das Sarma,
 Phys. Rev. Lett {\bf 101}, 125301 (2008).}
\bibitem{machholm2004PRA}{M. Machholm, A. Nicolin, C. J. Pethick, and H. Smith, Phys. Rev. A {\bf 69}, 043604 (2004).}











\end{thebibliography}



\end{document}